\documentclass[final,3p,times,twocolumn,authoryear]{elsarticle}

\usepackage{amssymb}
\usepackage{amsmath}
\usepackage{booktabs}
\usepackage{subcaption}
\usepackage[colorlinks=true, linkcolor=blue, citecolor=blue, urlcolor=blue]{hyperref}
\usepackage{xcolor}
\usepackage{ulem}
\usepackage{tabularx}

\journal{Nuclear Physics B}

\begin{document}

\begin{frontmatter}



\title{Hadronic origin of gamma rays and neutrinos from blazars: Multi-messenger implications and observational constraints} 


\author[inst1]{Rodrigo Sasse\corref{cor1}}
\ead{rodrigo.sasse1@uel.br} 

\author[inst2,inst1,inst3,inst4,inst5]{R. C. dos Anjos}

\cortext[cor1]{Corresponding author}


\address[inst1]{Programa de Pós-graduação em Física \& Departamento de Física, Universidade Estadual de Londrina (UEL), Rodovia Celso Garcia Cid Km 380, Londrina, 86057-970, Paraná, Brazil}

\address[inst2]{Centro de Artes, Humanidades e Tecnologia, Universidade Federal de São Carlos (UFSCar), R. Dr. Eduardo Nielsen, 420, Jardim Congonhas, 15030-070 São José do Rio Preto, SP, Brazil}


\address[inst3]{Programa de Pós-Graduação em Física e Astronomia, Universidade Tecnológica Federal do Paraná (UTFPR), Av. Sete de Setembro, 3165, Curitiba, 80230-901, Paraná, Brazil}

\address[inst4]{Programa de Pós-Graduação em Física Aplicada, Universidade Federal da Integração Latino-Americana, Av. Tarquínio Joslin dos Santos, 1000, Foz do Iguaçu, 85867-670, Paraná, Brazil}

\address[inst5]{Núcleo de Astrofísica e Cosmologia (Cosmo-Ufes), Departamento de Física, Universidade Federal do Espírito Santo, Av. Fernando Ferrari, 514, Vitória, 29075-910, Espírito Santo, Brazil}

\begin{abstract}
We investigate whether cosmogenic gamma rays and neutrinos from ultra-high-energy cosmic rays (UHECRs) produced in hadronically active blazars can significantly contribute to the observed spectral energy distribution (SED), and examine the potential of the Cherenkov Telescope Array Observatory (CTAO) to detect and discriminate this component. For four nearest neutrino bright blazars modeled by \citet{xrodrigues}, specifically AP Librae, TXS 1700+685, PKS 2326$-$502, and PKS 2345$-$16, we constrain the UHECR luminosity by requiring that the total emission does not overshoot available multi-wavelength data, assuming acceleration in the optically thin large scale jet, propagate UHECRs through realistic Galactic and extragalactic magnetic fields, and forecast CTAO performance. We find that cosmogenic gamma rays contribute up to about 7.5\% of the total gamma-ray emission at the highest energies ($10^{20}$\,eV), and thus remain subdominant in the SED. In contrast, the associated neutrino emission, at approximately $10^{44}$ to $10^{46}$\,erg\,s$^{-1}$, offers a robust propagation signature of hadronic acceleration. Our simulations of 50\,hour CTAO exposures show that the observatory can accurately reconstruct the intrinsic gamma ray spectrum, enabling strong constraints on the hadronic origin and propagation of UHECRs.  These results demonstrate the power of combining CTAO gamma-ray measurements with neutrino observations to probe blazar hadronic processes.
\end{abstract}




\end{frontmatter}



\section{Introduction} \label{sec:intro}

Blazars are a subclass of active galactic nuclei (AGNs) characterized by relativistic jets oriented close to our line of sight. These jets are among the most powerful particle accelerators in the Universe, capable of producing radiation spanning the entire electromagnetic spectrum, from radio waves to very-high-energy (VHE) gamma rays~\citep{2019Galax...7...20B, 2019NewAR..8701541H, 2022Galax..10..105S}. Their broadband emission has traditionally been interpreted within leptonic frameworks, where synchrotron and inverse Compton processes dominate the observed spectra. However, a growing body of multi-messenger evidence suggests that blazars are also sources of high-energy neutrinos, implying a hadronic or leptohadronic contribution to their emission. This connection has motivated renewed interest in hadronic models, which provide a more comprehensive and detailed description of particle acceleration and radiation mechanisms in relativistic jets.~\citep{Buson_2022, 10.1093/mnras/stad1467, 2018Sci...361.1378I, 2019NatAs...3...88G, 2019MNRAS.483L..12C, 2018ApJ...865..124M, 2022Sci...378..538I, 10.1093/mnras/stx3354, xrodrigues}.

Blazars have also been proposed as potential sources of ultra-high-energy cosmic rays (UHECR), with particles reaching energies up to $\sim10^{20}$ $\mathrm{eV}$. In these systems, relativistic jets launched by supermassive black holes can accelerate charged particles to nearly the speed of light through diffusive shock (first order Fermi) or stochastic (second order Fermi) acceleration processes. The accelerated particles may subsequently interact with the intense photon fields within the jet or with ambient matter near the central engine, producing high-energy gamma rays and neutrinos through hadronic channels such as $p\gamma$ or $pp$ interactions~\citep{doi:10.1142/9789811282645, 2012ApJ...749...63M, 2013ApJ...768...54B, 2019ApJ...874L..29R, 2020NewAR..8901543M}.

In these extreme environments, hadronic processes play a crucial role in shaping the multi-messenger signatures of blazars. When UHECR, primarily protons, interact with the dense radiation fields produced within the jet, photohadronic reactions ($p\gamma$ interactions) occur, leading to the production of mesons. Neutral pions ($\pi^0$) rapidly decay into pairs of high-energy gamma rays, contributing to the observed gamma-ray emission from these sources. In contrast, charged pions ($\pi^{\pm}$) decay into muons and subsequently into electrons, positrons, and neutrinos, establishing a direct link between the high-energy photon and neutrino fluxes~\citep{ANCHORDOQUI20191, 2014PhRvD..90b3007M, 2019MNRAS.483L..12C, 2015MNRAS.448.2412P, 2019NatAs...3...88G}.

UHECR can also interact with ambient matter within or around the relativistic jet, leading to inelastic proton–proton ($pp$) collisions. These interactions likewise result in pion production and, consequently, secondary gamma rays and neutrinos, through decay chains similar to those initiated in $p\gamma$ processes~\citep{refId0, 2006PhRvD..74c4018K}. The detection of high-energy neutrinos from sources such as the blazar TXS~0506+056 has provided compelling evidence for hadronic activity in these systems, marking a milestone in multi-messenger astrophysics~\citep{2018Sci...361..147I, IceCube:2023oua, 2018ApJ...863L..10A}. Joint observations of gamma rays and neutrinos from the same astrophysical source not only support the hypothesis that blazars contribute to the UHECR population, but also offer valuable insights into the extreme physical conditions governing their jets such as magnetic field strength, photon density, and the baryonic composition of the outflow.

Recent observations, including the detection of the ultra-high-energy (UHE) cosmic neutrino KM3~230213A by the KM3NeT observatory, have provided valuable information about the physics of extreme astrophysical environments \citep{km3net_2025, Adriani_2025}. While initial assessments considered a cosmogenic origin for this event, recent studies demonstrate that such an interpretation is not straightforward. Specifically, attributing this neutrino to cosmogenic processes strictly requires the presence of a subdominant proton component at the highest energies, because heavy nuclei are significantly less efficient at producing high-energy neutrinos via photo-pion production. Reconciling the observed neutrino energy and flux with existing cosmic ray data thus requires contributions from distant and powerful UHECR accelerators, alongside this crucial proton fraction. Consequently, a cosmogenic origin cannot be assumed as the default scenario without specific compositional constraints, leaving direct astrophysical production mechanisms as highly competitive alternatives. Ultimately, these complexities in pinpointing the exact origin of single UHE neutrino events emphasize the critical importance of multi-messenger observations for constraining the nature, composition, and cosmological evolution of UHECR sources, as well as for improving our understanding of their propagation over cosmological distances \citep{2020PhRvL.125l1104A}.

This work investigates the hadronic processes responsible for the production of gamma rays and neutrinos resulting from the propagation of UHECRs that may originate in blazars. The study examines both current theoretical frameworks and the observational evidence that support these scenarios. Understanding these mechanisms is essential for elucidating the nature of the most extreme cosmic particle accelerators. The primary goal of this study is to evaluate the contribution of UHECR propagation to the generation of cosmogenic particles, namely gamma rays and neutrinos, and to quantify how the resulting secondary flux influences the SED of each blazar. Furthermore, we assess the capabilities of the CTAO in detecting these gamma-ray signatures and analyze the observability of individual sources within its energy range~\citep{2019scta.book.....C}.

Unveiling the origin of UHECRs and establishing their link to blazars requires a multi-messenger framework that integrates intrinsic emission processes with cosmic propagation effects. This study aims to provide a self-consistent test of the hadronic scenario in selected blazars by quantifying the interdependence between different messengers. To this end, the paper is organized around three central goals: deriving the neutrino luminosity \( L_{\nu} \) and the corresponding ultra-high-energy cosmic ray luminosity \( L_{\mathrm{UHECR}} \) for our source sample, using the muon-neutrino flux predictions from the lepto-hadronic model of \citet{xrodrigues} as the benchmark for hadronic activity; propagating the injected UHECRs through detailed three-dimensional models of the extragalactic (EGMF) and Galactic (GMF) magnetic fields to compute the total survival fraction \( \xi_{\mathrm{total}} \) and the resulting cosmogenic gamma-ray and neutrino fluxes; and testing the capability of the CTAO to reconstruct the intrinsic and cosmogenic gamma-ray spectra, and to constrain the hadronic acceleration scenario for a 50 hours exposure. The paper is structured as follows. Section \ref{sec:Methodology} describes the methodology used to achieve these goals, including source selection, UHECR propagation simulations, and CTAO observation simulations. Section \ref{results} presents the results of these simulations, and Section \ref{discussion} discusses the implications and concludes.

\section{Methodology} \label{sec:Methodology}

\begin{figure}[ht!]
    \centering 
        \includegraphics[width=0.49\textwidth]{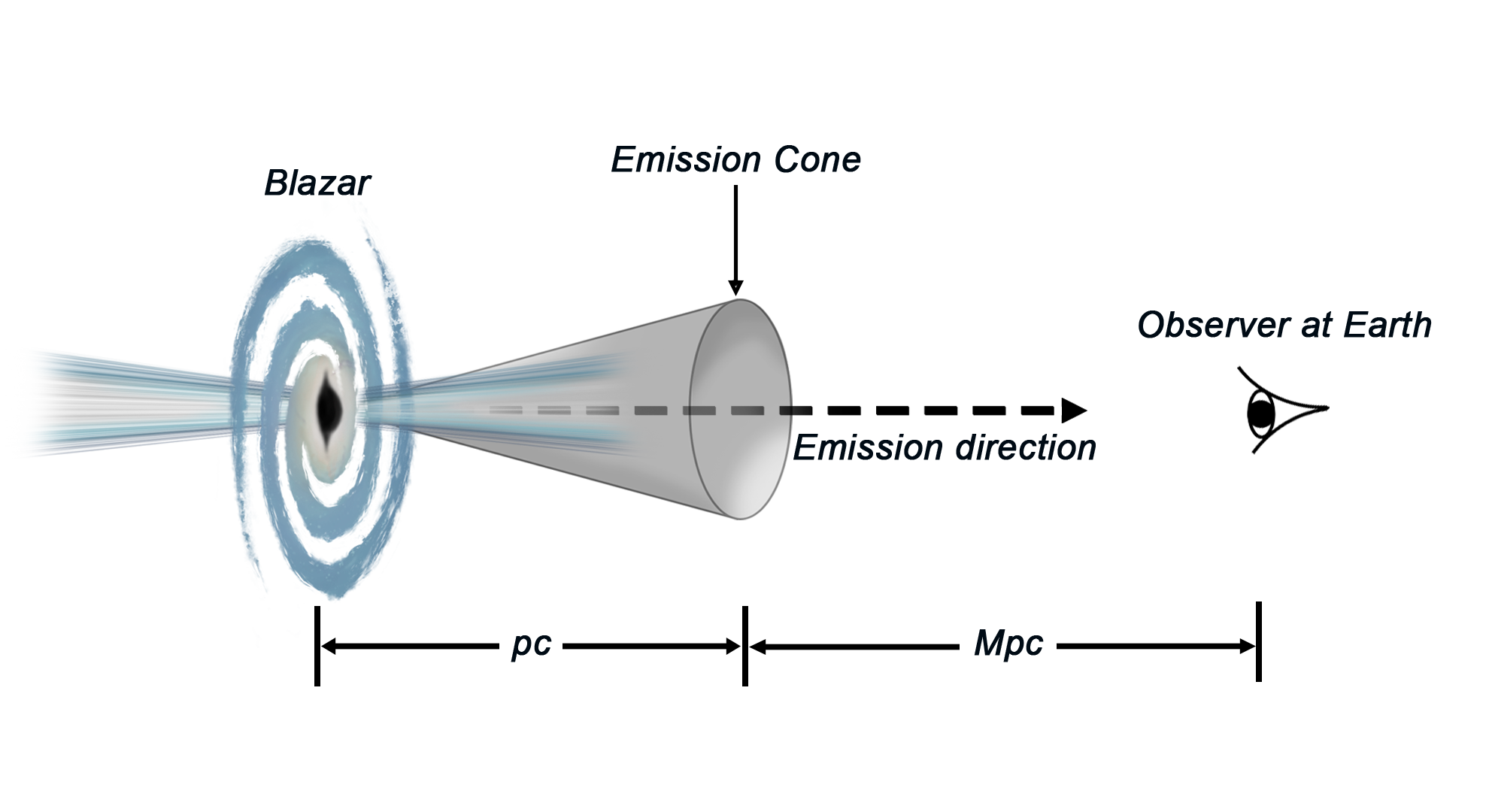}
    \caption{Schematic representation of the simulation setup adopted for each blazar, placed at its corresponding distance from the observer on Earth. The emission cone indicates the jet direction and the line of sight along which the ultra high energy particles are injected and propagated. The opening angle of the emission cone, which defines the simulated beam geometry, varies between $0^{\circ}$ and $1^{\circ}$ depending on the source distance.}
    \label{fig:config}
\end{figure}

This section presents the numerical simulation setup and the methodology adopted to connect the properties of UHECR sources with the resulting secondary gamma ray fluxes. The analysis involves the simulation of UHECR propagation and the computation of their associated cosmogenic secondary fluxes as a function of the distance to each source considered in this study. The simulations were performed with the \texttt{CRPropa3} code~\citep{AlvesBatista_2022}, a widely used framework within the UHECR research community that provides a flexible modular architecture for modeling different cosmic ray propagation scenarios. Figure~\ref{fig:config} shows the configuration adopted to simulate various source cases, aiming to reproduce the directional emission of UHECR originating from individual blazars.

For this study, we performed numerical simulations using two distinct models of the extragalactic magnetic field (EGMF). The first model adopts a Kolmogorov type turbulent magnetic field spectrum that follows a power law distribution across a range of turbulence scales defined by $L_{\mathrm{min}}$ and $L_{\mathrm{max}}$, which represent the smallest and largest coherence lengths, respectively. The simulations were carried out in a four-dimensional framework, requiring the specification of several physical parameters and the definition of the simulation geometry, as illustrated in Figure~\ref{fig:config}. 

The simulations employ the \textit{ObserverSurface} module of \texttt{CRPropa3}, in which the observer is represented as a spherical surface (detector) with a fixed radius, $R_{\mathrm{obs}} = 0.1~\mathrm{Mpc}$, centered at the propagation distance corresponding to each selected source. The coherence scales of the magnetic field are chosen to be comparable to the observer size, with $L_{\mathrm{min}} = 60~\mathrm{kpc}$ and $L_{\mathrm{max}} = 365~\mathrm{kpc}$. The spectral index of the Kolmogorov turbulence spectrum used in this configuration is $k = 5/3$. 

The second model corresponds to the \textit{astrophysical} scenario developed by \citet{10.1093/mnras/stx3354}. This model is based on a magnetohydrodynamic (MHD) representation of the EGMF obtained from a large scale cosmological simulation performed with the \texttt{ENZO} code. The simulation realistically reproduces the Local Universe through a constrained approach that generates structures analogous to known galaxy clusters such as Virgo and Coma. In this framework, the magnetic field arises self-consistently from energetic events occurring within galaxies and clusters, which magnetize the intergalactic medium and produce a detailed three-dimensional map of the magnetic field structure in the nearby Universe. 

This magnetic field map is implemented as the background environment in \texttt{CRPropa3} to simulate the trajectories of UHECR as they propagate through the intergalactic medium. The effects of both magnetic field models on the propagation of the injected particles are analyzed and discussed in Section~\ref{EGMF_effects}.

\subsection{Source Selection}
\label{sourceselection}

The source selection criteria adopted in this work were based on the approach developed by \citet{xrodrigues}, who analyzed a sample of 324 blazars from the Candidate Gamma-ray Blazar Survey (CGRaBS) catalogue~\citep{Healey_2008}. For each source, they employed a self-consistent numerical model known as leptohadronic, which simulates the radiation emitted by a population of relativistic electrons (leptonic component) and protons (hadronic component) accelerated within the emitting region of a blazar jet. The model predictions were fitted to the multiwavelength observational data of each blazar, covering the spectral range from radio to gamma rays. The fitting procedure was performed in three stages: initially using a purely leptonic model, then progressively introducing the hadronic contribution to evaluate whether it improved the agreement with the observed data.

For most of the sources (218 blazars, corresponding to 66\% of the sample), a purely leptonic model adequately reproduced the observed emission across multiple wavelengths. However, for 106 blazars (33\% of the sample), the inclusion of a hadronic component significantly improved the fit, particularly in the X-ray band. The analysis revealed an important trend: blazars with higher gamma-ray luminosities (GeV energies) tend to exhibit lower baryonic loading, although they are, on average, more efficient neutrino emitters. This behavior is consistent with the constraints derived from the IceCube stacking analyses~\citep{2017ApJ...835..151A, 2020PhRvL.124e1103A, 2022PhRvD.106b2005A}.

From the subset of 106 blazars whose spectral fits were improved by a hadronic component, we first identified those with the highest predicted neutrino fluxes. From this high-flux sample, we then selected the four nearest sources for detailed analysis. The final sample includes four sources: AP~Librae (z = 0.05), a nearby low-frequency-peaked BL Lac object characterized by an unusually broad high-energy component extending up to TeV energies~\citep{2015A&A...573A..31H, 2016A&A...588A.110Z}; TXS~1700+685 (z = 0.30), a flat-spectrum radio quasar with strong GeV emission detected by \textit{Fermi}-LAT and associated with a significant predicted neutrino flux~\citep{2022MNRAS.515.4675B}; PKS~2326$-$502 (z = 0.52), a bright southern-hemisphere FSRQ exhibiting pronounced gamma-ray variability and favorable conditions for photohadronic interactions~\citep{2017ApJ...835..182D}; and PKS~2345$-$16 (z = 0.58), a powerful FSRQ that displays intense flaring activity in both the X-ray and gamma-ray bands~\citep{2019ApJ...870...28F}. The comoving distance of each source was incorporated into the propagation simulations. In the following section, we describe how the results obtained from the leptohadronic modeling were integrated into our numerical simulations of UHECR propagation and secondary particle production.

\subsection{Numerical Methodology}
\label{nummethodology}

To investigate the contribution of cosmogenic gamma-ray and neutrino fluxes resulting from the propagation of UHECR, we analyzed each selected blazar individually. Our analysis builds upon the leptohadronic modeling framework developed by \citet{xrodrigues}, from which the predicted neutrino fluxes were obtained. 

The following sections detail the procedure to determine the maximum allowed UHECR luminosity, \(L_{\mathrm{UHECR}}\), constrained by multi-wavelength observations. Sections \ref{EGMF_effects} and \ref{GMF_effects} then quantify the fraction of these CR that survive propagation through the extragalactic (\(\xi_{\mathrm{EGMF}}\)) and Galactic (\(\xi_{\mathrm{GMF}}\)) magnetic fields to reach the observer. These transmission factors are essential for calculating the final, attenuated cosmogenic flux.

We assume that UHECRs are accelerated in a more
optically thin region of the blazar, such as the large-scale jet,
where protons can reach ultra-high energies and escape efficiently
without undergoing catastrophic interactions that would overproduce
in source electromagnetic cascades. For the propagation simulations,
we inject a primary proton spectrum following a power law
$dN/dE \propto E^{-\alpha}$ with spectral index $\alpha = 2.0$,
over an energy range from $E_{\mathrm{min}} = 10^{16}$~eV to a maximum
cutoff $E_{\mathrm{cut}} = 10^{20}$~eV. The UHECR luminosity
$L_{\mathrm{UHECR}}$ is then fixed by requiring that the total
combined flux (intrinsic + cosmogenic) does not overshoot the
observational upper limits provided by \textit{Fermi}-LAT and
H.E.S.S. data. This procedure yields a physically robust and
observationally grounded upper limit for the expected cosmogenic
signal, in line with the methodology of recent data driven
multi-messenger studies \citep{Das_2020, Das_2022}. The neutrino luminosity, $L_{\nu}$, for each blazar was then computed using
\begin{equation}
    L_{\nu} = 2\pi D_{\text{cm}}^2 \lbrack 1 - \cos(\theta_{\mathrm{jet}})\rbrack 
    \int_{\mathrm{E_{\nu,min}}}^{\mathrm{E_{\nu,max}}} 
    E_{\nu} \frac{dN}{dE_{\nu} dA dt}\, dE_{\nu},
    \label{eq:L_nu}
\end{equation}
where $D_{\text{cm}}$ is the comoving distance of the source, and the term $(1 - \cos\theta_{\mathrm{jet}})$ accounts for the emission confined within the jet opening angle, as defined in the simulation geometry. The total neutrino flux, $\frac{dN}{dE_{\nu} dA dt}$, was obtained from the leptohadronic modeling of \citet{xrodrigues}.

The geometric factor, $(1 - \cos(\theta_{\mathrm{jet}}))$ restricts the luminosity calculation to emission within the jet's solid angle, avoiding the assumption of isotropic emission from the source. The jet opening angle $\theta_{\mathrm{jet}}$ was estimated from the root-mean-square (RMS) deflection of cosmic-ray particles during propagation, following the approximation of \citet{Dermer_2009}:
\begin{equation}
    \Phi_{\mathrm{RMS}} \approx 4^{\circ}
    \frac{10^{2}\ \mathrm{EeV}}{E/Z}
    \frac{B_{\mathrm{RMS}}}{10^{-9}\ \mathrm{G}}
    \sqrt{\frac{D}{100\ \mathrm{Mpc}}}
    \sqrt{\frac{L_{\mathrm{c}}}{1\ \mathrm{Mpc}}},
    \label{deflection}
\end{equation}
where $L_{\mathrm{c}}$ represents the magnetic field coherence length, set to 0.1~Mpc in this study. Assuming a proton with a maximum energy of $E_{\mathrm{max}} = 10^2$~EeV, the estimated deflection is approximately $ \approx 0.0057^{\circ}$ assuming the PKS 2345-16 comoving distance and $B_{RMS} = 10^{-14}$ G.

The cosmogenic gamma-ray luminosity is calculated from
\begin{equation}
    L_{\gamma}^{\mathrm{UHECR}} = 
    \frac{2\pi D_{\text{cm}}^2 \lbrack 1 - \cos(\theta_{\mathrm{jet}})\rbrack}{\xi_{\mathrm{total}}}
    \int_{\mathrm{E_{\gamma,min}}}^{\mathrm{E_{\gamma,max}}}
    E_{\gamma} \frac{dN}{dE_{\gamma} dA dt}\, dE_{\gamma},
    \label{eq:L_uhecr_gamma}
\end{equation}
where the integral gamma-ray flux is obtained from simulations at each source distance, and $\xi_{\mathrm{total}}$ accounts for the fraction of injected UHECR that reach the observer:
\begin{equation}
    \xi_{\mathrm{total}} = \xi_{\mathrm{EGMF}} \times \xi_{\mathrm{GMF}}.
    \label{eq:xi_total}
\end{equation}

This correction incorporates the effects of both extragalactic and galactic magnetic fields on UHECR propagation by weighting the total luminosity according to the fraction of particles that successfully reach the observer. In this way, the method incorporates magnetic field attenuation when relating the intrinsic UHECR luminosity to the observable cosmogenic flux. The results obtained for $\xi_{\mathrm{EGMF}}$ and $\xi_{\mathrm{GMF}}$ for each magnetic field model and source distance are discussed in Sections~\ref{EGMF_effects} and~\ref{GMF_effects}.

Finally, the integral gamma-ray flux derived from Equation~\ref{eq:L_uhecr_gamma} must be normalized using the luminosity obtained from the leptohadronic modeling. The normalization condition is defined as
\begin{equation}
    L_{\gamma}^{\mathrm{UHECR}} = L_{\mathrm{lephad}}^{\mathrm{UHECR}}.
\end{equation}
The comparison between the neutrino and gamma-ray luminosities will be presented and discussed in Section~\ref{results}.

\subsection{Extragalactic Magnetic Field}
\label{EGMF_effects}

\begin{figure}[ht!]
    \centering 
        \includegraphics[width=0.49\textwidth]{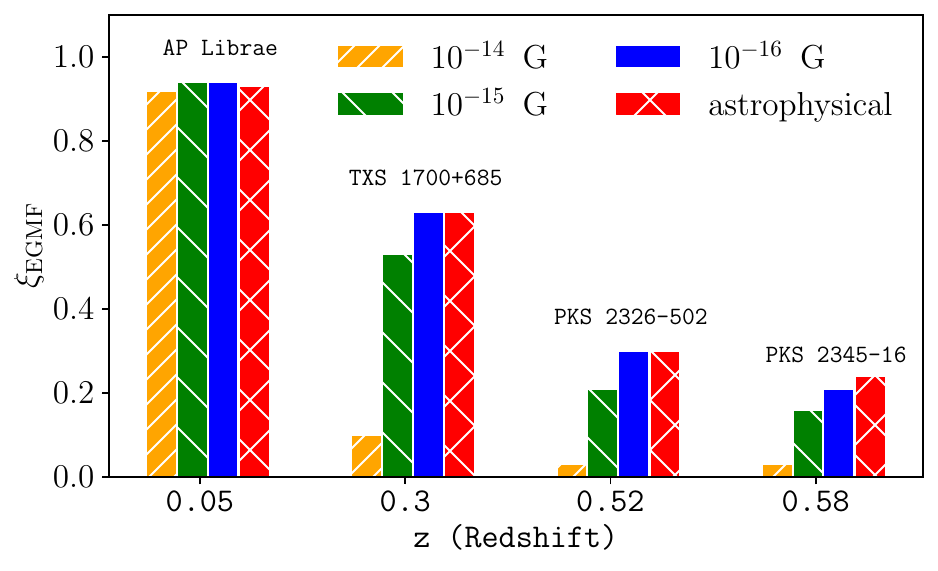}
    \caption{Fraction of injected protons, $\xi_{\mathrm{EGMF}}$, that successfully reach the observer as a function of the comoving distance of the source. The simulations assume a cutoff energy of $E_{\mathrm{cut}} = Z \times 10^{20}$~eV. Results are shown for the Kolmogorov turbulent model with three field strengths ($B_{\mathrm{RMS}} = 10^{-14}$, $10^{-15}$, and $10^{-16}$~G) and for the \textit{astrophysical} MHD model.}
    \label{fig:egmf}
\end{figure}

Figure~\ref{fig:egmf} presents the impact of the EGMF on the propagation of ultra-high-energy protons with a cutoff energy of $10^{20}$~eV, from each source to an observer located at Earth. Two different magnetic-field scenarios were considered. The first corresponds to a Kolmogorov type turbulent field, evaluated for three root-mean-square field strengths ($B_{\mathrm{RMS}}$). The second is the \textit{astrophysical} MHD model, which provides a more detailed three-dimensional field structure based on cosmological simulations. The Figure~\ref{fig:egmf} shows the fraction of injected protons, $\xi_{\mathrm{EGMF}}$, that successfully reach the observer for each of the four blazars considered, plotted as a function of redshift ($z$).

The results exhibit two clear behaviors. First, the fraction of protons arriving at the observer decreases systematically with increasing source distance. For the nearest source, AP~Librae ($z = 0.05$), the arrival fraction exceeds $90\%$ in almost all configurations. For more distant sources such as TXS~1700+685 ($z = 0.3$), PKS~2326$-$502 ($z = 0.52$), and PKS~2345$-$16 ($z = 0.58$), this fraction decreases sharply, falling below $30\%$ in most cases. This behavior reflects the cumulative deflection of charged particles as they traverse larger cosmic volumes permeated by magnetic fields.

Within the Kolmogorov turbulence scenario, the field strength ($B_{\mathrm{RMS}}$) has a decisive influence on the arrival fraction. Stronger magnetic fields produce larger deflections and therefore smaller arrival fractions. For example, for TXS~1700+685, the strongest field considered ($10^{-14}$~G) allows only about $10\%$ of the injected protons to reach the observer, whereas the weakest field ($10^{-16}$~G) permits more than $60\%$ to arrive.

The \textit{astrophysical} MHD model, represented by the red bars in the figure, shows a distinct pattern. For distant sources, it predicts higher arrival fractions than the Kolmogorov models with stronger fields ($10^{-14}$~G and $10^{-15}$~G), yielding results similar to those of the Kolmogorov configuration with the weakest field ($10^{-16}$~G). This outcome indicates that the magnetic-field structure obtained in the MHD simulation, which reflects a realistic large-scale distribution of galaxies and voids, is generally less effective at deflecting UHE protons. The weaker magnetization of intercluster and void regions in this model reduces the overall deflection probability, leading to higher arrival fractions at Earth.

\subsection{Galactic Magnetic Field}
\label{GMF_effects}

In our analysis, we also consider the deflection experienced by injected protons due to the GMF. Since the GMF can be several orders of magnitude stronger than the EGMF, its influence on the propagation of UHECR is significant. The average deflection angle of a proton in the Milky Way can be approximated by

\begin{equation}
    \theta_{\mathrm{def,MW}} \approx \frac{0.9^{\circ}}{\sin b}
    \left(\frac{60\ \mathrm{EeV}}{E / Z}\right)
    \left(\frac{B}{10^{-9}\ \mathrm{G}}\right)
    \left(\frac{h_{\mathrm{disk}}}{1\ \mathrm{kpc}}\right),
\end{equation}
where $b$ is the Galactic latitude of the source, and $h_{\mathrm{disk}}$ represents the vertical scale height of the Galactic disk. This formulation follows the approach presented by \citet{Dermer_2009, Das_2020, Das_2025}. The resulting deflection is incorporated into Equation~\ref{eq:xi_total} through the factor $\xi_{\mathrm{GMF}}$, which quantifies the fraction of particles successfully reaching the observer after accounting for the effects of the GMF.

The deflection was estimated using the JF12 model \citep{Jansson_2012, Unger}, a widely adopted reference framework for studying cosmic-ray propagation within the Milky Way. The JF12 model includes both large-scale and turbulent components of the magnetic field, constrained by extensive observational data such as Faraday rotation measures from extragalactic sources and polarized and total synchrotron emission maps obtained by the \textit{Wilkinson Microwave Anisotropy Probe} (WMAP) satellite. Its detailed structure enables a realistic representation of the magnetic environment that UHECR encounter within the Galaxy.

To evaluate the GMF effect, we backtracked anti-nuclei from the observer’s position to a radius of 20~kpc using the JF12 model. The simulated particles follow a power-law energy spectrum ($dN/dE \propto E^{-2.0}$) in the range from 1 to 100~EeV. A Gaussian angular smearing of $\sigma = 0.107$~rad was applied to include uncertainties associated with both magnetic-field fluctuations and the experimental angular resolution. Following the procedure adopted by the Pierre Auger Observatory, circular regions with a radius of $15^{\circ}$ (corresponding to a 95\% confidence level) were defined around each candidate source for the analysis \citep{2019JCAP...10..022A}.

\begin{figure*}[ht]
    \centering 
        \includegraphics[width=0.99\textwidth]{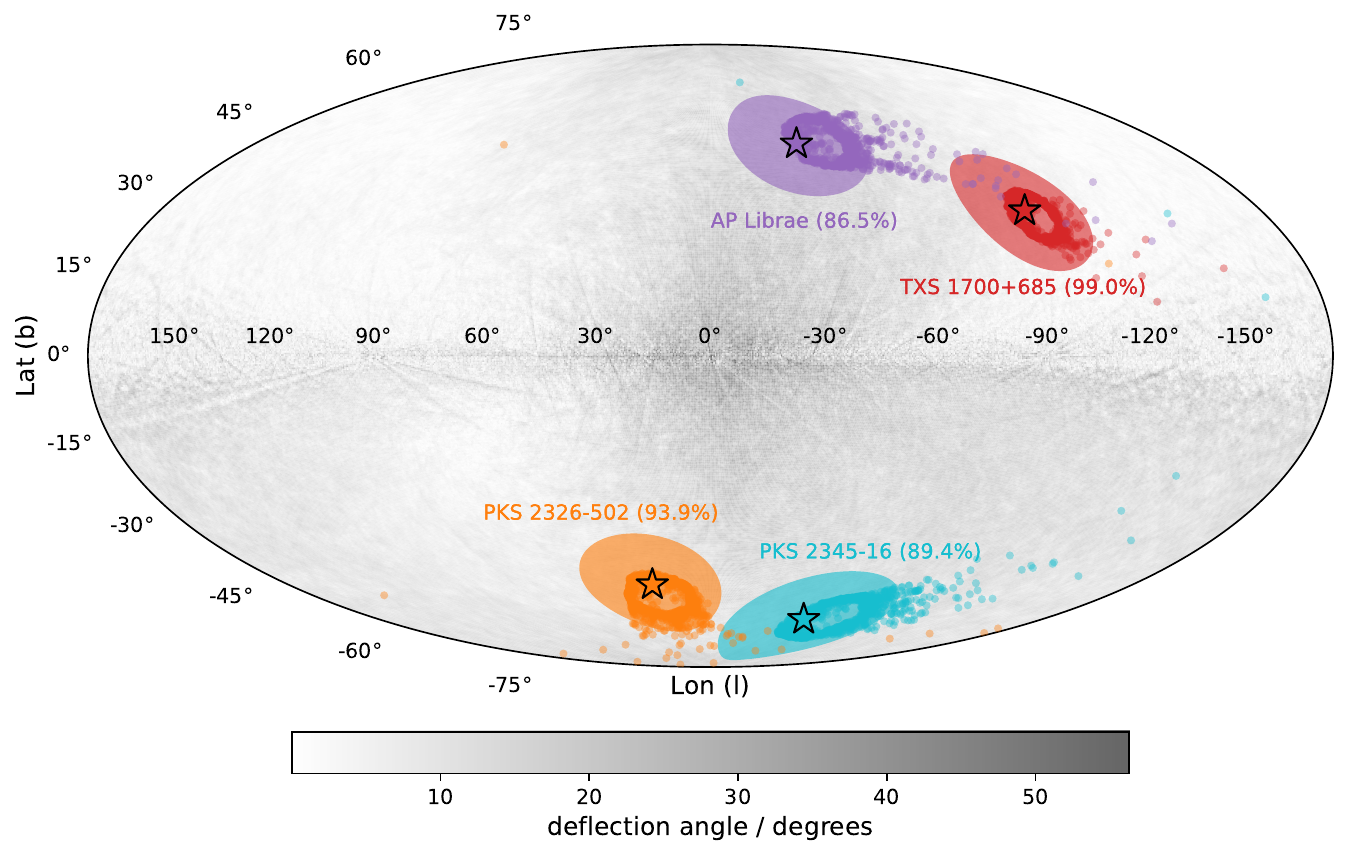}
    \caption{All-sky map in Hammer projection showing the average angular deflection of protons produced by both the regular and turbulent components of the galactic magnetic field, modeled with the JF12 framework. The map is displayed in galactic coordinates, with the Galactic center located at $(\ell, b) = (0^{\circ}, 0^{\circ})$. Star symbols mark the positions of the candidate blazar sources, each identified by name. The colored contours represent the 95\% confidence regions derived from the backtracked arrival directions of simulated particles. The percentage given next to each source indicates the fraction of simulated protons that reach the observer within the corresponding confidence region.}
    \label{fig:gmf}
\end{figure*}

Figure~\ref{fig:gmf} shows the all-sky distribution of the average deflection angles experienced by ultra-high-energy protons as they traverse the Galactic magnetic field. The map exhibits a clear anisotropy, with the largest deflections concentrated toward the Galactic Center, where the magnetic-field structure is highly intricate and the field strength reaches its maximum \citep{Jansson_2012}. In contrast, smaller deflections are observed at high Galactic latitudes, where the magnetic field is weaker and more ordered. The skymap shows how the geometry and intensity of the Galactic magnetic field modulate the propagation of UHECR depending on their incoming direction.

For each candidate blazar, the backtracked arrival directions of simulated antiprotons (shown as dots) form a distinct spatial probability distribution on the celestial sphere. The colored contours correspond to the 95\% confidence regions that enclose the majority of simulated trajectories. The fraction of particles arriving within these regions, represented by $\xi_{\mathrm{GMF}}$, quantifies the effective transmission probability of UHECR through the Galaxy for each source. This factor is subsequently applied as a weight to the simulated cosmic ray flux when evaluating the effective gamma-ray luminosity ($L_{\gamma}^{\mathrm{UHECR}}$). In this way, the combined influence of both extragalactic and galactic magnetic deflections is consistently incorporated, ensuring that the predicted cosmogenic fluxes represent conservative, physically realistic estimates.

\subsection{CTAO Performance}

\label{CTAO_performance}

To evaluate the detectability and characterize the spectral properties of the blazars AP~Librae, TXS~1700+685, PKS~2326-502, and PKS~2345-16 with the future CTAO, we performed detailed observation simulations. The cosmogenic gamma-ray component derived in the previous sections adds a small but non-negligible high-energy tail to the intrinsic leptohadronic spectra. We therefore simulate CTAO observations using the combined (intrinsic + cosmogenic) spectral models as input. This approach allows us to assess whether CTAO can: detect these sources in the multi-TeV range and accurately reconstruct their intrinsic spectral parameters, and place empirical constraints on the allowed level of cosmogenic contribution, which in turn limits the plausible UHECR luminosity $L_{\mathrm{UHECR}}$ of each blazar. The simulations were performed using the open-source software package \texttt{Gammapy}~\citep{gammapy:2023}. Our analysis follows the methodology described by \citet{Costa2024}, further refined by \citet{Sousa_2025} and \citet{Sasse_2025}.

The intrinsic spectral shape of each blazar was modeled using the best-fit parameters provided by the \textit{Fermi}-LAT 4FGL catalog. For all sources, we adopted a Log-Parabola spectral model (\texttt{logParabolaSpectralModel}), with the parameters $\alpha$ and $\beta$ taken directly from the catalog. The only exception is AP~Librae, for which the spectral model derived by H.E.S.S.~\citep{2015A&A...573A..31H} was used; this model is also a Log-Parabola function and ensures consistency with the TeV observational data. The choice between the CTA-North and CTA-South arrays for each source was determined from an annual visibility study. Each simulation considered a total observation time of 50~hours considering as a cumulative observation accumulated over multiple cycles of CTAO operations, using a standard “wobble” pointing strategy with a $0.5^{\circ}$ offset from the nominal source position.

The Instrument Response Functions (IRFs) were selected to match the optimal observing conditions for each source. For every configuration, IRFs corresponding to the most favorable zenith angle were applied and optimized for a reference observation time of 50~hours. The visibility time, IRF setup, and spectral parameters adopted for each blazar are summarized in Table~\ref{tab:table1}.

\begin{table*}[htbp]
\centering
\small
\setlength{\tabcolsep}{4pt}
\caption{Input and fitted spectral parameters used in the CTAO
simulations for the four candidate blazars. The \textit{Input Model}
corresponds to the intrinsic spectral shape adopted from the
\textit{Fermi}-LAT 4FGL catalog (and H.E.S.S. for AP~Librae),
described by a Log-Parabola function
$dN/dE = N_0\,(E/E_{\mathrm{ref}})^{-(\alpha + \beta \log(E/E_{\mathrm{ref}}))}$,
with amplitudes $N_0$ in units of $\mathrm{cm^{-2}\,s^{-1}\,MeV^{-1}}$.
The \textit{Fitted Model} lists the parameters reconstructed from
50~h CTAO simulated observations. \textit{Annual Visibility} is
reported for the array used for each source (CTA-South or CTA-North)
at zenith angles of $20^{\circ}/40^{\circ}/60^{\circ}$; ``---'' means
the source is not visible at that zenith from that array.}
\label{tab:table1}
\begin{tabular}{lcccc}
\toprule
 & \textbf{AP~Librae} & \textbf{TXS~1700+685} & \textbf{PKS~2326$-$502} & \textbf{PKS~2345$-$16} \\
 & (4FGL J1517.7$-$2422) & (4FGL J1700.0+6830) & (4FGL J2329.3$-$4955) & (4FGL J2348.0$-$1630) \\
\midrule
\multicolumn{5}{l}{\textit{Input Model} \,(4FGL\,$+$\,H.E.S.S. for AP~Librae)} \\
$N_0$            & $2.205\times10^{-13}$            & $(1.33\pm0.02)\times10^{-11}$  & $(4.6\pm0.06)\times10^{-11}$    & $(8.84\pm0.19)\times10^{-12}$ \\
$E_{\mathrm{ref}}$ & $5.48$~GeV                      & $700.67$~MeV                    & $560.29$~MeV                    & $872.50$~MeV \\
$\alpha$         & $2.239$                          & $2.249\pm0.02$                  & $2.177\pm0.01$                  & $2.215\pm0.02$ \\
$\beta$          & $0.051$                          & $0.097\pm0.01$                  & $0.117\pm0.01$                  & $0.070\pm0.01$ \\
\midrule
\multicolumn{5}{l}{\textit{Fitted Model} \,(CTAO, 50~h)} \\
$N_0$            & $(2.07\pm0.11)\times10^{-13}$    & $(1.47\pm0.35)\times10^{-11}$   & $(1.07\pm0.31)\times10^{-10}$   & $(1.06\pm0.11)\times10^{-11}$ \\
$E_{\mathrm{ref}}$ & $5.48$~GeV                      & $700$~MeV                       & $560$~MeV                       & $872$~MeV \\
$\alpha$         & $2.210\pm0.04$                   & $2.299\pm0.09$                  & $2.427\pm0.09$                  & $2.29\pm0.04$ \\
$\beta$          & $0.056\pm0.01$                   & $0.092\pm0.009$                 & $0.099\pm0.01$                  & $0.063\pm0.01$ \\
\midrule
\multicolumn{5}{l}{\textit{Annual Visibility} (h) at zenith $20^{\circ}/40^{\circ}/60^{\circ}$} \\
CTA-South        & $538.0\,/\,543.5\,/\,555.0$      & ---                             & $513.5\,/\,859.5\,/\,761.0$     & $606.5\,/\,537.0\,/\,538.5$ \\
CTA-North        & $\text{---}\,/\,\text{---}\,/\,1151.0$ & $0\,/\,1282.0\,/\,1374.5$ & ---                             & $\text{---}\,/\,541.0\,/\,807.0$ \\
\bottomrule
\end{tabular}
\end{table*}

A one-dimensional spectral analysis was then performed for each blazar. The region of interest (ON region) was defined as a circular aperture with a radius of $0.5^{\circ}$ centered on the source position. The background was estimated using the reflected regions method (\texttt{on-off}), with the total acceptance area of the OFF regions set to be five times larger than that of the ON region.

Each simulated dataset incorporates instrument exposure, background modeling, and energy dispersion (\texttt{edisp}). Two energy axes were defined: a reconstructed energy range from 30~GeV to 300~TeV (divided into five bins per decade) and a true energy range from 3~GeV to 500~TeV, the latter chosen to minimize boundary effects due to energy dispersion. To assess statistical robustness, 100 independent observation realizations (\texttt{n\_obs}~=~100) were generated for each source, enabling the evaluation of the variance in fitted parameters.

The simulated data, including Poisson statistical fluctuations, were fitted with the corresponding intrinsic spectral models to reconstruct the best-fit parameters. From these fits, spectral energy distribution (SED) points were derived for each source, dividing the energy range into five bins per decade between 30~GeV and 300~TeV.

\section{Results}\label{results}

In this section, we present and interpret the results obtained for the four selected blazars, focusing on their potential role as accelerators of UHECR. We discuss the corresponding cosmogenic gamma-ray and neutrino fluxes derived from our simulations, as well as the detectability of these sources with the future CTAO.

\subsection{Cosmogenic Flux Contribution and Luminosities}


Figure~\ref{fig:mwl_results} summarizes the predicted spectral energy distributions (SEDs) for AP~Librae (\ref{fig:4a}), TXS~1700+685 (\ref{fig:4b}), PKS~2326$-$502 (\ref{fig:4c}), and PKS~2345$-$16 (\ref{fig:4d}), combining the results from the leptohadronic model of~\citet{xrodrigues} with the cosmogenic components obtained in this work.
The figure displays the gamma-ray and neutrino spectra associated with both the intrinsic emission and the propagation-induced (cosmogenic) contributions, for the maximum acceleration energy of, $E_{\mathrm{max}} = 10^{20}$~eV. Observational data from \textit{Fermi}-LAT (green triangles) and H.E.S.S. (orange squares) are included for comparison, along with the CTAO differential sensitivity curve for 50~h exposure.

It is important to note the spatial distribution of these secondary components. The cosmogenic neutrinos, produced in interactions with background photons during propagation, are effectively isotropic at the angular scale of a single source. Therefore, their predicted flux should be interpreted as a contribution to the diffuse astrophysical neutrino background rather than a strictly point-like signal. For the cosmogenic gamma rays, the situation is more complex. While the electromagnetic cascades from UHECR interactions can be broadened by intermediate magnetic fields, our adopted magnetic field configurations (Section~\ref{EGMF_effects}) and the associated deflections are relatively weak. The curves shown in Fig.~\ref{fig:mwl_results} thus represent an upper limit to the point-like contribution, assuming the secondary emission remains within the line-of-sight cone defined by the jet opening angle. In more detailed scenarios with stronger magnetic fields, a significant fraction of the cosmogenic gamma-ray power would be redistributed as extended or even diffuse emission. Consequently, adding the cosmogenic component directly to the point-source SED provides an optimistic but observationally constrained estimate of its potential detectability, as any angular broadening would only further dilute its surface brightness.

As demonstrated by \cite{xrodrigues}, the lephadronic models reproduce the observed SEDs across a broad energy range, from GeV to TeV energies. The additional cosmogenic gamma-ray component, shown by the red curves, emerges at the highest energies ($E \gtrsim 10^{7}$~eV) but remains below the dominant intrinsic component. For all sources, the predicted cosmogenic neutrino flux (dark blue lines) reaches levels that are observationally relevant at EeV energies. While current detectors like IceCube lack the effective volume to resolve this specific component for individual sources, this flux falls within the projected differential discovery potential of next generation observatories. In this sense, future facilities such as IceCube-Gen2 and GRAND will possess the necessary effective area to potentially detect these high-energy neutrino signatures over their planned operational lifetimes. Based on these results, we quantified the contribution of the cosmogenic gamma-ray component to the total emission of each blazar. Table~\ref{tab:table2} summarizes these contributions and the associated luminosities. For the scenario with a maximum acceleration energy of $E_{\mathrm{max}} = 10^{20}$~eV, the cosmogenic flux contributes up to 7.5\% of the total gamma-ray emission, depending on the source.

From the combined models, we derived the corresponding neutrino (\(L_{\nu}\)) and UHECR (\(L_{\mathrm{UHECR}}^{\mathrm{lephad}}\)) luminosities for each source (see Table~\ref{tab:table2}). The cosmogenic fractions reported in Table~\ref{tab:table2} follow directly from the total transmission factors \(\xi_{\mathrm{total}}\) derived in Sections~\ref{EGMF_effects}-\ref{GMF_effects}, thereby encapsulating the combined effect of extragalactic and Galactic magnetic fields on the link between the intrinsic leptohadronic source power and the observable secondary flux. The estimated UHECR luminosities lie in the range of \(10^{44}\)--\(10^{46}\)~erg~s\(^{-1}\), consistent with the expected energetics of powerful astrophysical accelerators such as blazar jets. The inferred neutrino luminosities provide a direct test of the hadronic scenario, as neutrinos can reach Earth without significant absorption during propagation. While UHECR are deflected by magnetic fields and lose energy through interactions with background radiation, and gamma rays are attenuated via pair production on the extragalactic background light, neutrinos travel along straight trajectories, preserving both their energy and directional information. This property makes them the most reliable messengers for confirming hadronic acceleration in distant active galaxies.

The detection of these cosmogenic neutrinos will, therefore, be a critical step in validating the hadronic nature of blazar jets. Future large-volume detectors, such as IceCube-Gen2~\citep{Aartsen_2021}, GRAND~200k~\citep{neto2023giantradioarrayneutrino}, and KM3NeT~\citep{km3net_2025}, are expected to reach the sensitivity required to detect this diffuse cosmogenic flux.

\begin{figure*}[htbp]
    \centering 
    \begin{subfigure}[b]{0.9\textwidth}
        \centering
        \includegraphics[width=\textwidth]{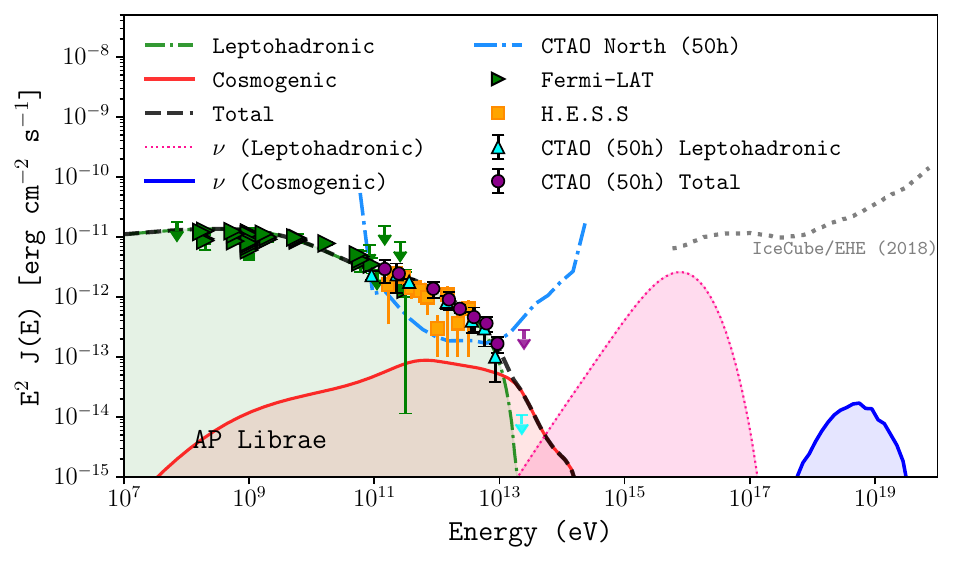}
        \caption{AP Librae SED showing the combined leptohadronic modeling, cosmogenic flux predictions, and simulated CTAO performance. \textit{Fermi}-LAT points are green triangles and H.E.S.S. are orange squares. The dashed green line is the \citet{xrodrigues} leptohadronic model, and the pink line is its neutrino flux. The dashed black line is the Total model. Solid red and dark blue lines are the cosmogenic gamma-ray and neutrino fluxes ($E_{\mathrm{max}} = 10^{20}$~eV). The blue dash-dotted line is the 50~h CTAO sensitivity, with simulated flux points as cyan triangles/purple circles.}
        \label{fig:4a}
    \end{subfigure}
    
    \vspace{0.5cm}

    \begin{subfigure}[b]{0.9\textwidth}
        \centering
        \includegraphics[width=\textwidth]{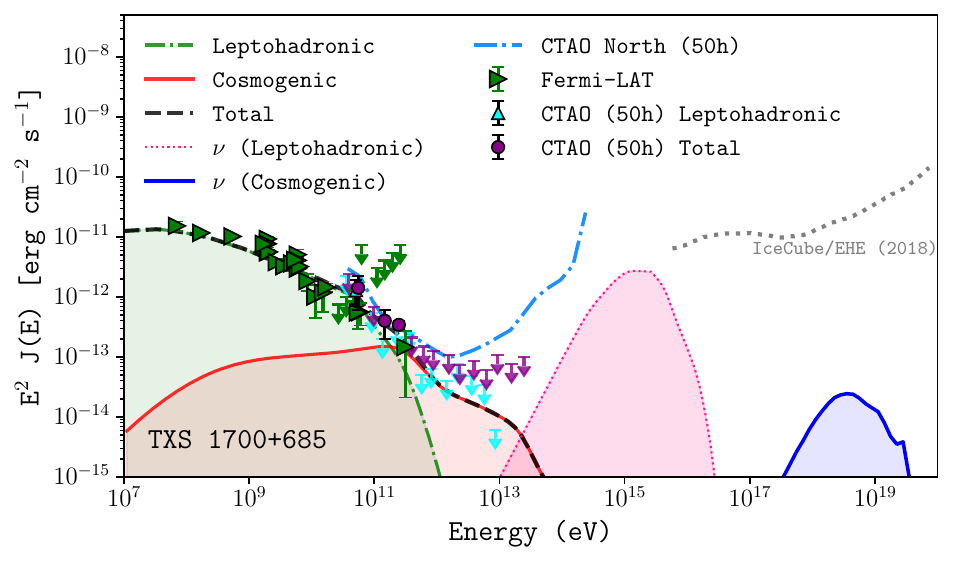}
        \caption{TXS 1700+685 Multi-messenger SED following the same curve and marker conventions as panel (a), with the 50~h CTAO differential sensitivity computed for this specific source declination.}
        \label{fig:4b}
    \end{subfigure}
\end{figure*}

\begin{figure*}[htbp]
    \ContinuedFloat
    \centering
    \begin{subfigure}[b]{0.9\textwidth}
        \centering
        \includegraphics[width=\textwidth]{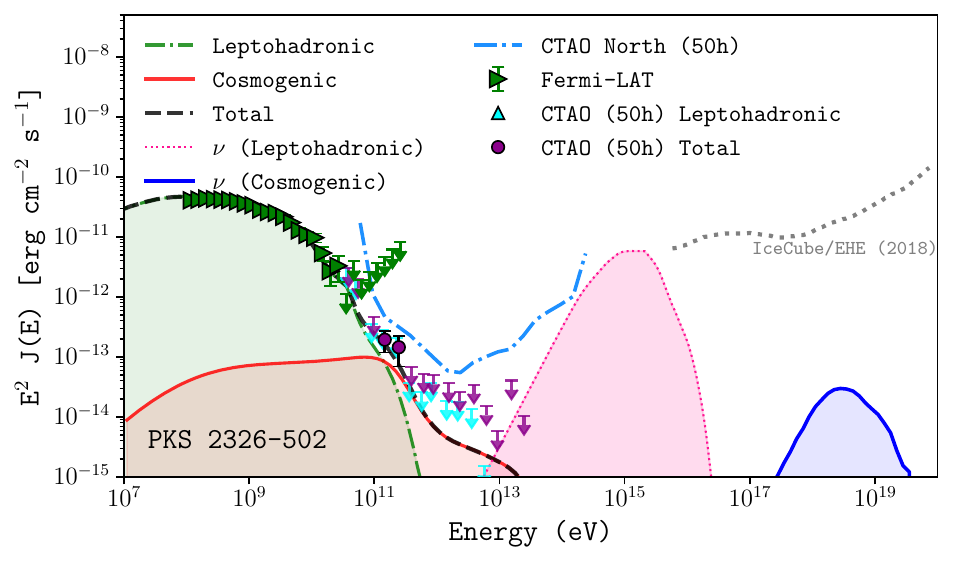}
        \caption{PKS 2326-502 Multi-messenger SED following the same curve and marker conventions as panel (a), with the 50~h CTAO differential sensitivity computed for this specific source declination.}
        \label{fig:4c}
    \end{subfigure}
    
    \vspace{0.5cm}

    \begin{subfigure}[b]{0.9\textwidth}
        \centering
        \includegraphics[width=\textwidth]{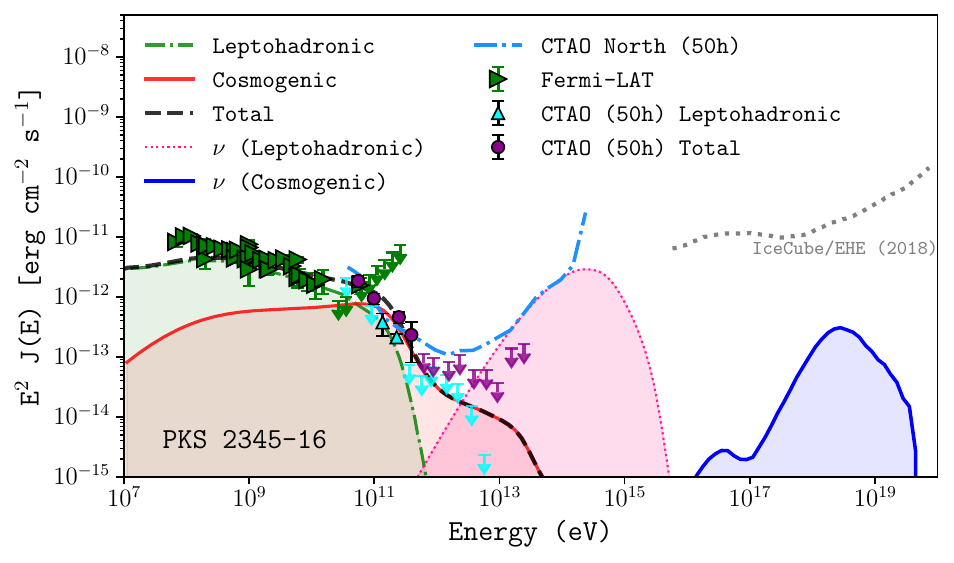}
        \caption{PKS 2345-16 Multi-messenger SED following the same curve and marker conventions as panel (a), with the 50~h CTAO differential sensitivity computed for this specific source declination.}
        \label{fig:4d}
    \end{subfigure}

    \caption{Multi-messenger SEDs for the four candidate blazars. Detailed descriptions of the theoretical models, multi-wavelength data, and simulated CTAO performance are provided in the individual panel captions.}
    \label{fig:mwl_results}
\end{figure*}

\begin{table*}[htbp]
\centering
\renewcommand{\arraystretch}{1.3}
\setlength{\tabcolsep}{12pt}
\caption{Summary of the cosmogenic gamma-ray contribution and the
corresponding neutrino and UHECR luminosities for the four candidate
blazars. The second column lists the fractional contribution of the
cosmogenic gamma-ray component to the total leptohadronic emission
for a maximum acceleration energy $E_{\mathrm{max}} = 10^{20}$~eV.
The third and fourth columns give the derived neutrino luminosity
$L_{\nu}$ and the UHECR luminosity inferred from the leptohadronic
model $L_{\mathrm{lephad}}^{\mathrm{UHECR}}$. All luminosities are in
units of $\mathrm{erg\,s^{-1}}$.}
\label{tab:table2}
\begin{tabular*}{\textwidth}{@{\extracolsep{\fill}}lccc}
\toprule
\textbf{Source} &
\textbf{$\gamma$-ray Cosmogenic} &
\textbf{Neutrino Luminosity} &
\textbf{UHECR Luminosity} \\
 &
\textbf{Coverage (\%)} &
\textbf{$L_{\nu}$ (erg\,s$^{-1}$)} &
\textbf{$L_{\mathrm{lephad}}^{\mathrm{UHECR}}$ (erg\,s$^{-1}$)} \\
\midrule
AP~Librae       & 0.5 & $7.9 \times 10^{43}$  & $4.21 \times 10^{44}$ \\
TXS~1700+685    & 2.0 & $6.14 \times 10^{44}$ & $3.27 \times 10^{45}$ \\
PKS~2326$-$502  & 0.3 & $6.1 \times 10^{45}$  & $3.25 \times 10^{46}$ \\
PKS~2345$-$16   & 7.5 & $3.80 \times 10^{44}$ & $2.00 \times 10^{45}$ \\
\bottomrule
\end{tabular*}
\end{table*}

\subsection{UHECR Propagation Effects and the Extragalactic Magnetic Field}

Although the calculated luminosities are physically consistent with the energetics of powerful blazar jets, establishing a direct association between observed UHECR and individual sources remains extremely challenging due to propagation effects. During their journey from the source to Earth, charged particles are deflected by both the extragalactic and galactic magnetic fields, which distort their arrival directions and reduce the number of particles that reach the observer within the line of sight.

Figure~\ref{fig:egmf} shows the dependence of the proton arrival fraction on distance and on the strength of the EGMF. The results demonstrate a clear trend: the fraction of protons reaching the observer decreases as the source redshift and the magnetic field strength increase. For the nearby blazar AP~Librae ($z=0.05$), more than ninety percent of the injected protons reach the observer in most scenarios. However, for more distant sources such as PKS~2345$-$16 ($z=0.58$), this fraction can drop below twenty percent, even for the weakest magnetic field considered ($B_{\mathrm{RMS}} = 10^{-16}$~G). This behavior shows the crucial role of magnetic deflections in shaping the observed CR anisotropy and explains the absence of clear directional correlations between known blazars and the arrival directions of UHECR.

When combined with the galactic magnetic field map shown in Figure~\ref{fig:gmf}, the overall picture suggests that deflections within the Galaxy dominate the final arrival direction of UHECR, while the EGMF primarily affects the total transmission fraction of particles from each source. Even under optimistic conditions, the cumulative influence of both magnetic environments severely limits the possibility of reconstructing the original trajectories of individual UHECR. These results reinforce the need for a multi-messenger approach to constrain the origin of UHECR, since neutral messengers are not subject to magnetic deflections.

\subsection{CTAO Performance for Blazars}
\label{ctao_blazars}

The simulations confirm the capability of CTAO to detect and precisely characterize the gamma-ray emission from the selected blazars after 50 hours of observation. Figure~\ref{fig:mwl_results} shows the reconstructed spectra for AP~Librae (\ref{fig:4a}), TXS~1700+685 (\ref{fig:4b}), PKS~2326$-$502 (\ref{fig:4c}), and PKS~2345$-$16 (\ref{fig:4d}), demonstrating that the fitted spectral energy distributions are consistent with the input models derived from the \textit{Fermi}-LAT 4FGL catalog and, where available, from H.E.S.S. observations. Table~\ref{tab:table1} summarizes the spectral parameters used as input and those reconstructed from the simulated CTAO data. For each source, the fitted parameters: amplitude, index $\alpha$, and curvature $\beta$, agree with the input values within statistical uncertainties, confirming that CTAO will accurately reproduce the intrinsic spectra and enable detailed studies of their emission mechanisms.

Given that the cosmogenic gamma-ray component contributes up to 7.5\% of the total flux (Table~\ref{tab:table2}), our simulations indicate that CTAO is primarily sensitive to the dominant intrinsic emission. Isolating the cosmogenic tail from the intrinsic spectrum with CTAO alone would be extremely challenging, emphasizing the need for complementary neutrino observations to constrain the hadronic acceleration scenario. Nevertheless, the precise spectral reconstruction achieved by CTAO will provide essential constraints on the overall emission model, against which any excess at the highest energies could be tested.

For three of the sources, the simulated fluxes based on the \textit{Fermi}-LAT and H.E.S.S.\ models lie below the nominal differential sensitivity curve of CTAO for a 50 hours exposure. Although this implies that detections with the conventional 5$\sigma$ significance would be challenging, the simulations indicate that CTAO can still measure these emissions at a confidence level of at least 3$\sigma$. These detections correspond to the flux points plotted in Figure~\ref{fig:mwl_results}; for energy bins not meeting this threshold, upper limits were derived. One hundred independent realizations were generated for each simulated observation to account for photon counting fluctuations. The CTAO flux points shown in Figure~\ref{fig:mwl_results} represent the expected distribution of measured outcomes, and the parameters listed in Table~\ref{tab:table1} correspond to the fit of the median dataset, ensuring that the reported values reflect the most probable result rather than a statistical outlier.

\begin{figure*}[t!] 
    \centering
    \begin{subfigure}{0.48\textwidth} 
        \includegraphics[width=\linewidth]{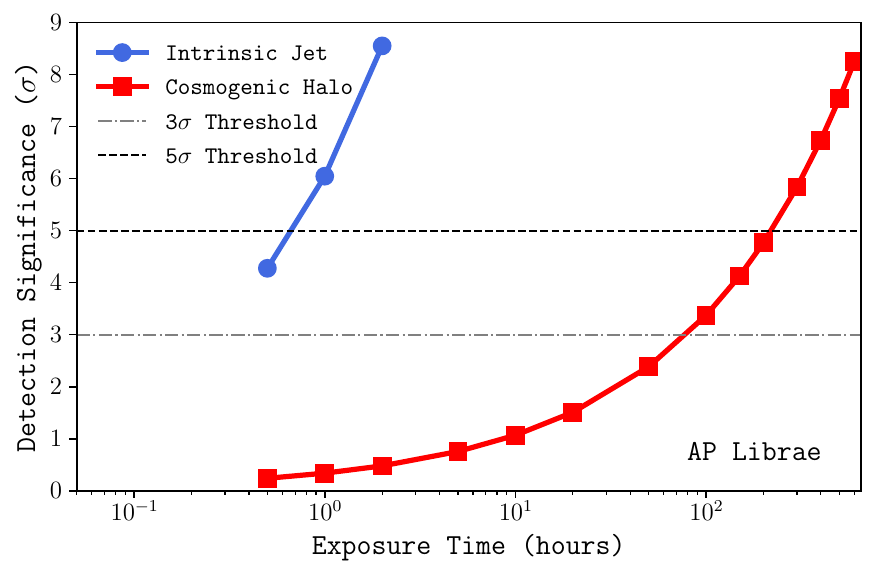}
        \caption{AP Librae}
        \label{fig:time_aplibrae}
    \end{subfigure}\hfill
    \begin{subfigure}{0.48\textwidth}
        \includegraphics[width=\linewidth]{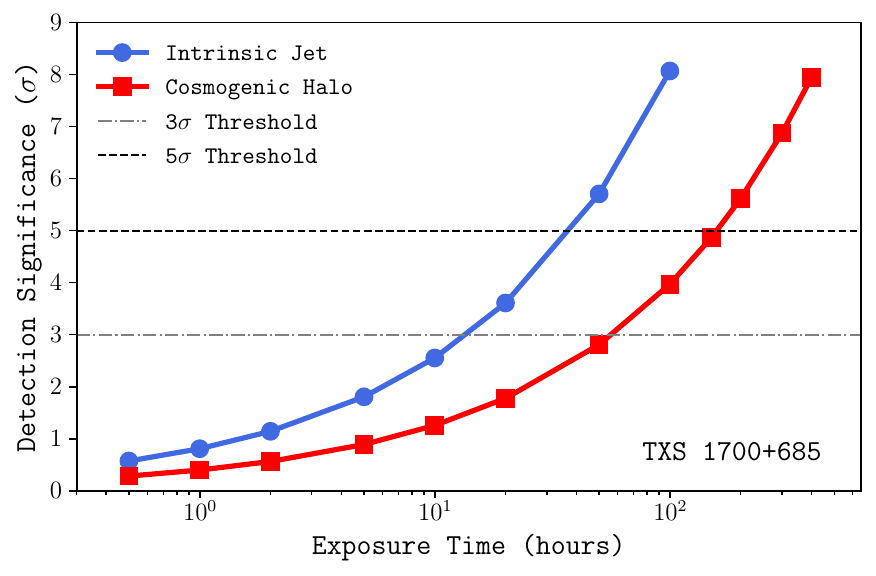}
        \caption{TXS~1700+685}
        \label{fig:time_txs1700}
    \end{subfigure}

    \vspace{0.3cm}

    \begin{subfigure}{0.48\textwidth}
        \includegraphics[width=\linewidth]{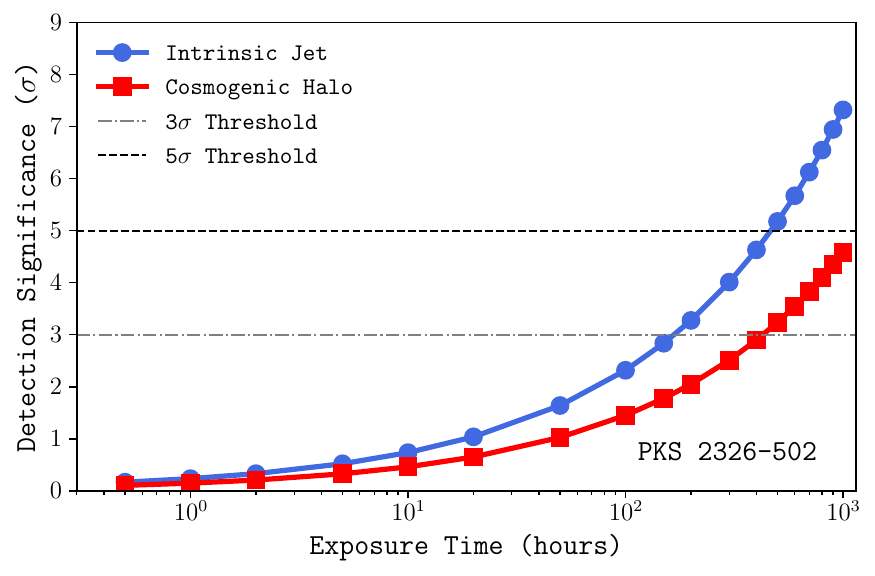}
        \caption{PKS~2326$-$502}
        \label{fig:time_pks2326}
    \end{subfigure}\hfill
    \begin{subfigure}{0.48\textwidth}
        \includegraphics[width=\linewidth]{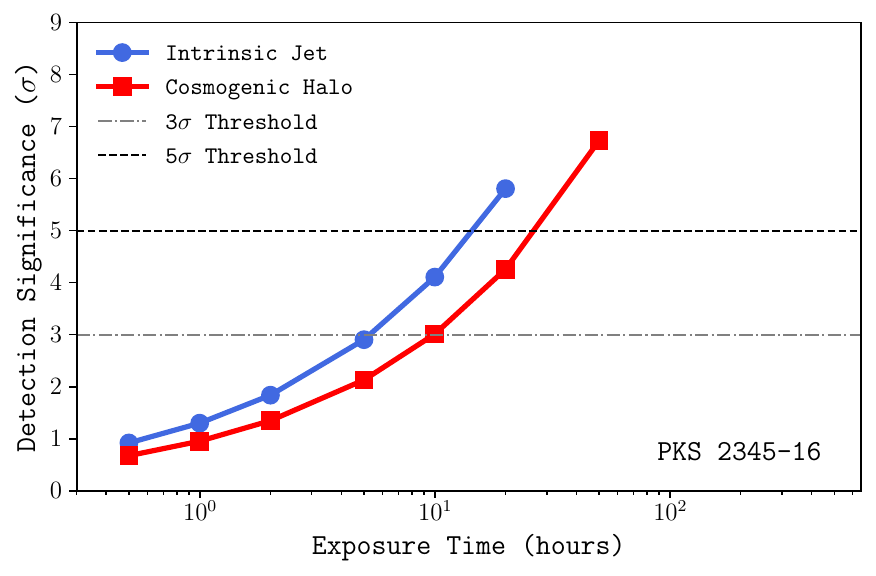}
        \caption{PKS~2345$-$16}
        \label{fig:time_pks2345}
    \end{subfigure}
    
    \caption{Detection significance ($\sigma$) as a function of exposure time for the four blazars analyzed, simulated with the baseline CTAO ``Alpha'' configuration. Blue circles show the point-like intrinsic jet and red squares the extended cosmogenic halo. Horizontal lines mark the $3\sigma$ (dash dotted grey) and $5\sigma$ (dashed black) detection thresholds. Exposure times range from 0.5 to 600~hours, extended to 1000~hours for PKS~2326$-$502 (c).}
    \label{fig:time_all_sources}
\end{figure*}

To quantify the observational prospects, we evaluated the integration time required for a detection. Figure~\ref{fig:time_all_sources} shows the detection significance ($\sigma$) as a function of exposure time, from 0.5 to 600~hours and up to 1000~hours for PKS~2326$-$502, for the CTAO ``Alpha'' configuration. The point-like intrinsic jet and the extended cosmogenic halo are evaluated independently, and the two components require very different exposures. The intrinsic jet emission reaches the $5\sigma$ discovery threshold within typical short to medium observational windows ($\sim 1$ to $50$~hours), which makes it an accessible target for early operations. The extended cosmogenic halo alone demands hundreds of hours of integration time. Exposures of this length are not feasible within a single observational cycle and must be understood as cumulative integration times built up over several years, most likely
within dedicated multi cycle observing programs or long-term Key Science
Projects~\citep{2019scta.book.....C}. The intrinsic point-like jet can therefore be readily probed, whereas detecting the extended cosmogenic component remains a long-term scientific objective.

The simulated observations thus demonstrate that CTAO will be capable of distinguishing between different emission and propagation scenarios using long exposure times, playing a decisive role in constraining the hadronic contribution to the gamma-ray emission of these blazars.

\section{Discussion and conclusion}\label{discussion}

This work provides a self consistent, multi-messenger test of the hadronic scenario for four neutrino bright blazars (AP Librae, TXS 1700+685, PKS 2326-502, and PKS 2345-16). We combined leptohadronic source modeling~\citep{xrodrigues} with \texttt{CRPropa3} propagation simulations and CTAO observation forecasts to achieve three interconnected goals: to constrain the UHECR luminosity by requiring that the total emission does not overshoot the available multi-wavelength data, assuming acceleration in the optically thin large-scale jet, to propagate these UHECRs through realistic Galactic and extragalactic magnetic fields to compute the resulting cosmogenic gamma-ray and neutrino fluxes, and to forecast CTAO's ability to probe these specific components. Our main conclusions are as follows.

We find that a cosmogenic gamma-ray component is consistently produced but remains subdominant. For a maximum proton acceleration energy of $E_{\max}=10^{20}$~eV, with the UHECR luminosity fixed at an optimistic upper limit that still does not overshoot the \textit{Fermi}-LAT and H.E.S.S. data, the cosmogenic component constitutes up to 7.5\% of the total gamma-ray emission. The predicted cosmogenic neutrino fluxes remain below the level that IceCube can resolve for individual sources, but they fall within the differential discovery potential of next generation
facilities, which makes them a robust tracer of hadronic activity.

Our propagation simulations elucidate the critical role of magnetic fields in modulating UHECR transport~\citep{Das_2020,Das_2025}. The fraction of injected protons reaching Earth decreases systematically with source distance and extragalactic magnetic field strength, dropping from over 90\% for the nearby blazar AP Librae to 30\% or less for more distant sources under typical turbulent field conditions. The Galactic magnetic field introduces additional, direction dependent deflections that are most pronounced toward the inner Galaxy~\citep{Das_2020,Das_2025,Dermer_2009}. This combined magnetic screening explains the longstanding challenge of establishing direct directional associations between UHECR events and individual blazars, reinforcing the necessity of neutral messengers for source identification.

Regarding observational prospects, our CTAO simulations separate two regimes. The point-like intrinsic gamma-ray emission is reliably reconstructed and constrained within typical short to medium observational windows, from about $1$ to $50$~hours for most targets. However, the extended cosmogenic gamma-ray tail requires integration times of hundreds to nearly a thousand hours. Its detection is therefore a long-term milestone that depends on data accumulated over several years, most plausibly within Key
Science Projects. The observatory's value during early operations lies in its precise spectral characterization of the dominant intrinsic emission, which, when combined with neutrino measurements from next-generation observatories like IceCube-Gen2 and KM3NeT, will enable powerful constraints on the hadronic scenario. Our results are contingent on several assumptions, including the maximum proton rigidity, the strength and structure of the extragalactic magnetic field, and the baryonic loading derived from the leptohadronic model~\citep[e.g.,][]{2017ApJ...835..151A, Aartsen_2021, 2022PhRvD.106b2005A}. Factors such as extragalactic background light attenuation and a mixed cosmic-ray composition could further modulate the predicted fluxes. Despite these caveats, the synergistic use of gamma-ray and neutrino data remains the most promising pathway to conclusively test hadronic acceleration in blazar jets and to elucidate the origin of ultra-high-energy cosmic rays~\citep{2019scta.book.....C, km3net_2025, IceCube:2023oua}.

\section*{Acknowledgments}
R.S. acknowledges financial support from the Coordenação de Aperfeiçoamento de Pessoal de Nível Superior – Brasil (CAPES) – Finance Code 001. R.S. and R.C.A. acknowledge the support of the NAPI “Fenômenos Extremos do Universo” of Fundação de Apoio à Ciência, Tecnologia e Inovação do Paraná. R.C.A. research is supported by CAPES/Alexander von Humboldt Program (88881.800216/2022-01), CNPq (308859/2025-1) and (4000045/2023-0), Araucária Foundation (698/2022) and (721/2022) and FAPESP (2021/01089-1). The authors acknowledge the AWS Cloud Credit/CNPq and the National Laboratory for Scientific Computing (LNCC/MCTI, Brazil) for providing HPC resources of the SDumont supercomputer, which have contributed to the research results reported in this paper. URL: https://sdumont.lncc.br. This work was developed with the support of the Laboratório Interinstitucional de e-Astronomia (LIneA).

\section*{Authors Contribution}
All authors contributed equally to this work.





\begin{thebibliography}{00}


\bibitem[Aab et al.(2019)]{2019JCAP...10..022A}
A. Aab et al. (The Pierre Auger Collaboration),
Probing the origin of ultra-high-energy cosmic rays with neutrinos in the EeV energy range using the Pierre Auger Observatory,
\textit{JCAP}, vol. 2019, no. 10, p. 022, October 2019.
doi: 10.1088/1475-7516/2019/10/022.

\bibitem[Aartsen et al.(2017)]{2017ApJ...835..151A}
M. G. Aartsen et al. (IceCube Collaboration),
All-sky Search for Time-integrated Neutrino Emission from Astrophysical Sources with 7 yr of IceCube Data,
\textit{ApJ}, vol. 835, no. 2, p. 151, February 2017.
doi: 10.3847/1538-4357/835/2/151.

\bibitem[Aartsen et al.(2020a)]{2020PhRvL.125l1104A}
M. G. Aartsen et al. (IceCube Collaboration),
Characteristics of the Diffuse Astrophysical Electron and Tau Neutrino Flux with Six Years of IceCube High Energy Cascade Data,
\textit{Phys. Rev. Lett.}, vol. 125, no. 12, p. 121104, September 2020.
doi: 10.1103/PhysRevLett.125.121104.

\bibitem[Aartsen et al.(2020b)]{2020PhRvL.124e1103A}
M. G. Aartsen et al. (IceCube Collaboration),
Time-Integrated Neutrino Source Searches with 10 Years of IceCube Data,
\textit{Phys. Rev. Lett.}, vol. 124, no. 5, p. 051103, February 2020.
doi: 10.1103/PhysRevLett.124.051103.

\bibitem[Aartsen et al.(2021)]{Aartsen_2021}
M. G. Aartsen et al. (IceCube Collaboration),
IceCube-Gen2: the window to the extreme Universe,
\textit{Journal of Physics G: Nuclear and Particle Physics}, vol. 48, no. 6, p. 060501, April 2021.
doi: 10.1088/1361-6471/abbd48.

\bibitem[Abbasi et al.(2022)]{2022PhRvD.106b2005A}
R. Abbasi et al. (IceCube Collaboration),
Search for neutrino emission from cores of active galactic nuclei,
\textit{Phys. Rev. D}, vol. 106, no. 2, p. 022005, July 2022.
doi: 10.1103/PhysRevD.106.022005.

\bibitem[Abbasi et al.(2023)]{IceCube:2023oua}
R. Abbasi et al. (IceCube Collaboration),
TXS 0506+056 with Updated IceCube Data,
\textit{PoS}, vol. ICRC2023, p. 1465, 2023.
doi: 10.22323/1.444.1465.

\bibitem[Adriani et al.(2025)]{Adriani_2025}
O. Adriani et al. (The KM3NeT Collaboration),
On the Potential Cosmogenic Origin of the Ultra-high-energy Event KM3-230213A,
\textit{The Astrophysical Journal Letters}, vol. 984, no. 2, p. L41, May 2025.
doi: 10.3847/2041-8213/adcc29.

\bibitem[Aharonian et al.(2023)]{aharonian}
F. Aharonian et al., Constraints on the Intergalactic Magnetic Field Using Fermi-LAT and H.E.S.S. Blazar Observations, \textit{The Astrophysical Journal Letters}, vol. 950, p. L16, 2023.
doi: 10.3847/2041-8213/acd777

\bibitem[Alves Batista et al.(2022)]{AlvesBatista_2022}
R. Alves Batista et al.,
CRPropa 3.2 — an advanced framework for high-energy particle propagation in extragalactic and galactic spaces,
\textit{Journal of Cosmology and Astroparticle Physics}, vol. 2022, no. 09, p. 035, September 2022.
doi: 10.1088/1475-7516/2022/09/035.

\bibitem[Anchordoqui(2019)]{ANCHORDOQUI20191}
L. A. Anchordoqui,
Ultra-high-energy cosmic rays,
\textit{Physics Reports}, vol. 801, pp. 1--93, 2019.
doi: 10.1016/j.physrep.2019.01.002.

\bibitem[Ansoldi et al.(2018)]{2018ApJ...863L..10A}
S. Ansoldi et al. (The MAGIC Collaboration),
The Blazar TXS 0506+056 Associated with a High-energy Neutrino: Insights into Extragalactic Jets and Cosmic-Ray Acceleration,
\textit{ApJL}, vol. 863, no. 1, p. L10, August 2018.
doi: 10.3847/2041-8213/aad083.

\bibitem[Banerjee et al.(2022)]{2022MNRAS.515.4675B}
A. Banerjee et al.,
Broadband spectro-temporal study on blazar TXS 1700+685,
\textit{MNRAS}, vol. 515, no. 4, pp. 4675--4684, October 2022.
doi: 10.1093/mnras/stac2068.

\bibitem[Böttcher et al.(2013)]{2013ApJ...768...54B}
M. Böttcher et al.,
Leptonic and Hadronic Modeling of Fermi-detected Blazars,
\textit{ApJ}, vol. 768, no. 1, p. 54, May 2013.
doi: 10.1088/0004-637X/768/1/54.

\bibitem[Böttcher(2019)]{2019Galax...7...20B}
M. Böttcher,
Progress in Multi-Wavelength and Multi-Messenger Observations of Blazars and Theoretical Challenges,
\textit{Galaxies}, vol. 7, no. 1, p. 20, January 2019.
doi: 10.3390/galaxies7010020.

\bibitem[Buson et al.(2022)]{Buson_2022}
S. Buson et al.,
Beginning a Journey Across the Universe: The Discovery of Extragalactic Neutrino Factories,
\textit{The Astrophysical Journal Letters}, vol. 933, no. 2, p. L43, July 2022.
doi: 10.3847/2041-8213/ac7d5b.

\bibitem[Cerruti et al.(2019)]{2019MNRAS.483L..12C}
M. Cerruti et al.,
Leptohadronic single-zone models for the electromagnetic and neutrino emission of TXS 0506+056,
\textit{MNRAS}, vol. 483, no. 1, pp. L12--L16, February 2019.
doi: 10.1093/mnrasl/sly210.

\bibitem[Cherenkov Telescope Array Consortium(2019)]{2019scta.book.....C}
Cherenkov Telescope Array Consortium,
\textit{Science with the Cherenkov Telescope Array},
World Scientific, 2019.
doi: 10.1142/10986.

\bibitem[Costa et al.(2024)]{Costa2024}
R. Jr. Costa et al.,
A gamma-ray study of galactic PeVatron candidates LHAASO J1825-1326 and LHAASO J1839-0545,
\textit{Journal of Cosmology and Astroparticle Physics}, vol. 2024, no. 07, p. 035, July 2024.
doi: 10.1088/1475-7516/2024/07/035.

\bibitem[Das et al.(2020)]{Das_2020}
S. Das, N. Gupta, and S. Razzaque,
Ultrahigh-energy Cosmic-Ray Interactions as the Origin of Very High-energy $\gamma$-Rays from BL Lacertae Objects,
\textit{The Astrophysical Journal}, vol. 889, no. 2, p. 149, February 2020.
doi: 10.3847/1538-4357/ab6131.

\bibitem[Das et al.(2022)]{Das_2022}
S. Das, S. Razzaque, and N. Gupta,
``Cosmogenic gamma-ray and neutrino fluxes from blazars associated with IceCube events'',
\textit{Astronomy \& Astrophysics}, vol. 658, p. L6, 2022.
doi: 10.1051/0004-6361/202142123

\bibitem[Das et al.(2025)]{Das_2025}
S. Das, S. Hazra, and N. Gupta,
Cosmic Clues from Amaterasu: Blazar-driven Ultrahigh-energy Cosmic Rays?,
\textit{The Astrophysical Journal Letters}, vol. 988, no. 1, p. L8, July 2025.
doi: 10.3847/2041-8213/ade99f.

\bibitem[de Mello Neto(2023)]{neto2023giantradioarrayneutrino}
J. R. T. de Mello Neto,
The Giant Radio Array for Neutrino Detection,
\textit{arXiv e-prints}, July 2023.
eprint: 2307.13638.

\bibitem[Dermer et al.(2009)]{Dermer_2009}
C. D. Dermer, S. Razzaque, J. D. Finke, and A. Atoyan,
Ultra-high-energy cosmic rays from black hole jets of radio galaxies,
\textit{New Journal of Physics}, vol. 11, no. 6, p. 065016, June 2009.
doi: 10.1088/1367-2630/11/6/065016.

\bibitem[Donath et al.(2023)]{gammapy:2023}
A. Donath et al.,
Gammapy: A Python package for gamma-ray astronomy,
\textit{A\&A}, vol. 678, p. A157, 2023.
doi: 10.1051/0004-6361/202346488.

\bibitem[Dutka et al.(2017)]{2017ApJ...835..182D}
M. S. Dutka et al.,
Multiband Observations of the Quasar PKS 2326-502 during Active and Quiescent Gamma-Ray States in 2010-2012,
\textit{ApJ}, vol. 835, no. 2, p. 182, February 2017.
doi: 10.3847/1538-4357/835/2/182.

\bibitem[Finke(2019)]{2019ApJ...870...28F}
J. D. Finke,
The Properties of Parsec-scale Blazar Jets,
\textit{ApJ}, vol. 870, no. 1, p. 28, January 2019.
doi: 10.3847/1538-4357/aaf00c.

\bibitem[Gao et al.(2019)]{2019NatAs...3...88G}
S. Gao, A. Fedynitch, W. Winter, and M. Pohl,
Modelling the coincident observation of a high-energy neutrino and a bright blazar flare,
\textit{Nature Astronomy}, vol. 3, pp. 88--92, January 2019.
doi: 10.1038/s41550-018-0610-1.

\bibitem[Hackstein et al.(2018)]{10.1093/mnras/stx3354}
S. Hackstein, F. Vazza, M. Brüggen, J. G. Sorce, and S. Gottlöber,
Simulations of ultra-high energy cosmic rays in the local Universe and the origin of cosmic magnetic fields,
\textit{Monthly Notices of the Royal Astronomical Society}, vol. 475, no. 2, pp. 2519--2529, January 2018.
doi: 10.1093/mnras/stx3354.

\bibitem[Healey et al.(2008)]{Healey_2008}
S. E. Healey et al.,
CGRaBS: An All-Sky Survey of Gamma-Ray Blazar Candidates,
\textit{The Astrophysical Journal Supplement Series}, vol. 175, no. 1, p. 97, March 2008.
doi: 10.1086/523302.

\bibitem[H.E.S.S. Collaboration(2015a)]{2015A&A...573A..31H}
H.E.S.S. Collaboration,
The high-energy $\gamma$-ray emission of AP Librae,
\textit{A\&A}, vol. 573, p. A31, January 2015.
doi: 10.1051/0004-6361/201321436.

\bibitem[Hovatta and Lindfors(2019)]{2019NewAR..8701541H}
T. Hovatta and E. Lindfors,
Relativistic Jets of Blazars,
\textit{NewAR}, vol. 87, p. 101541, December 2019.
doi: 10.1016/j.newar.2020.101541.

\bibitem[IceCube Collaboration(2018b)]{2018Sci...361.1378I}
IceCube Collaboration,
Multimessenger observations of a flaring blazar coincident with high-energy neutrino IceCube-170922A,
\textit{Science}, vol. 361, no. 6398, p. eaat1378, July 2018.
doi: 10.1126/science.aat1378.

\bibitem[IceCube Collaboration(2018c)]{2018Sci...361..147I}
IceCube Collaboration,
Neutrino emission from the direction of the blazar TXS 0506+056 prior to the IceCube-170922A alert,
\textit{Science}, vol. 361, no. 6398, pp. 147--151, July 2018.
doi: 10.1126/science.aat2890.

\bibitem[IceCube Collaboration(2022)]{2022Sci...378..538I}
R. Abbasi et al. (IceCube Collaboration),
Evidence for neutrino emission from the nearby active galaxy NGC 1068,
\textit{Science}, vol. 378, no. 6619, pp. 538--543, November 2022.
doi: 10.1126/science.abg3395.

\bibitem[Jansson and Farrar(2012)]{Jansson_2012}
R. Jansson and G. R. Farrar,
A New Model of the Galactic Magnetic Field,
\textit{The Astrophysical Journal}, vol. 757, no. 1, p. 14, August 2012.
doi: 10.1088/0004-637X/757/1/14.

\bibitem[Kelner et al.(2006)]{2006PhRvD..74c4018K}
S. R. Kelner, F. A. Aharonian, and V. V. Bugayov,
Energy spectra of gamma rays, electrons, and neutrinos produced at proton-proton interactions in the very high energy regime,
\textit{Phys. Rev. D}, vol. 74, no. 3, p. 034018, August 2006.
doi: 10.1103/PhysRevD.74.034018.

\bibitem[KM3NeT Collaboration(2025a)]{km3net_2025}
The KM3NeT Collaboration,
Observation of an Ultra-High-Energy Cosmic Neutrino with KM3NeT,
\textit{Nature}, vol. 638, pp. 376--382, 2025.
doi: 10.1038/s41586-024-08543-1.

\bibitem[Matthews et al.(2020)]{2020NewAR..8901543M}
J. H. Matthews, A. R. Bell, and K. M. Blundell,
Particle acceleration in astrophysical jets,
\textit{NewAR}, vol. 89, p. 101543, September 2020.
doi: 10.1016/j.newar.2020.101543.

\bibitem[Murase et al.(2012)]{2012ApJ...749...63M}
K. Murase et al.,
Blazars as Ultra-high-energy Cosmic-ray Sources: Implications for TeV Gamma-Ray Observations,
\textit{ApJ}, vol. 749, no. 1, p. 63, April 2012.
doi: 10.1088/0004-637X/749/1/63.

\bibitem[Murase et al.(2014)]{2014PhRvD..90b3007M}
K. Murase, Y. Inoue, and C. D. Dermer,
Diffuse neutrino intensity from the inner jets of active galactic nuclei: Impacts of external photon fields and the blazar sequence,
\textit{Phys. Rev. D}, vol. 90, no. 2, p. 023007, July 2014.
doi: 10.1103/PhysRevD.90.023007.

\bibitem[Murase et al.(2018)]{2018ApJ...865..124M}
K. Murase, F. Oikonomou, and M. Petropoulou,
Blazar Flares as an Origin of High-energy Cosmic Neutrinos?,
\textit{ApJ}, vol. 865, no. 2, p. 124, October 2018.
doi: 10.3847/1538-4357/aada00.

\bibitem[Murase and Stecker(2023)]{doi:10.1142/9789811282645}
K. Murase and F. W. Stecker,
High-Energy Neutrinos from Active Galactic Nuclei,
\textit{The Encyclopedia of Cosmology}, Chapter 10, pp. 483--540, 2023.
doi: 10.1142/9789811282645\_0010.

\bibitem[Petropoulou et al.(2015)]{2015MNRAS.448.2412P}
M. Petropoulou et al.,
Photohadronic origin of $\gamma$-ray BL Lac emission: implications for IceCube neutrinos,
\textit{MNRAS}, vol. 448, no. 3, pp. 2412--2429, April 2015.
doi: 10.1093/mnras/stv179.

\bibitem[Plavin et al.(2023)]{10.1093/mnras/stad1467}
A. V. Plavin, Y. Y. Kovalev, Yu. A. Kovalev, and S. V. Troitsky,
Growing evidence for high-energy neutrinos originating in radio blazars,
\textit{Monthly Notices of the Royal Astronomical Society}, vol. 523, no. 2, pp. 1799--1808, May 2023.
doi: 10.1093/mnras/stad1467.

\bibitem[Rodrigues et al.(2019)]{2019ApJ...874L..29R}
X. Rodrigues et al.,
Leptohadronic Blazar Models Applied to the 2014-2015 Flare of TXS 0506+056,
\textit{ApJL}, vol. 874, no. 2, p. L29, April 2019.
doi: 10.3847/2041-8213/ab1267.

\bibitem[Rodrigues et al.(2024)]{xrodrigues}
X. Rodrigues, V. S. Paliya, S. Garrappa, A. Omeliukh, A. Franckowiak, and W. Winter,
Leptohadronic multi-messenger modeling of 324 gamma-ray blazars,
\textit{A\&A}, vol. 681, p. A119, 2024.
doi: 10.1051/0004-6361/202347540.

\bibitem[Sasse et al.(2025)]{Sasse_2025}
R. Sasse et al.,
Blazars Jets and Prospects for TeV-PeV Neutrinos and Gamma Rays Through Cosmic-Ray Interactions,
\textit{Brazilian Journal of Physics}, vol. 55, no. 2, January 2025.
doi: 10.1007/s13538-024-01689-3.

\bibitem[Sol and Zech(2022)]{2022Galax..10..105S}
H. Sol and A. Zech,
Blazars at Very High Energies: Emission Modelling,
\textit{Galaxies}, vol. 10, no. 6, p. 105, November 2022.
doi: 10.3390/galaxies10060105.

\bibitem[Sousa et al.(2025)]{Sousa_2025}
M. F. Sousa et al.,
Prospects for Gamma-Ray Emission from Magnetar Regions in CTAO Observations,
\textit{The Astrophysical Journal}, vol. 979, no. 1, p. 23, January 2025.
doi: 10.3847/1538-4357/ad9b23.

\bibitem[Tchernin et al.(2013)]{refId0}
C. Tchernin et al.,
An exploration of hadronic interactions in blazars using IceCube,
\textit{A\&A}, vol. 555, p. A70, 2013.
doi: 10.1051/0004-6361/201220508.

\bibitem[Unger and Farrar(2024)]{Unger}
M. Unger and G. R. Farrar,
The Coherent Magnetic Field of the Milky Way,
\textit{The Astrophysical Journal}, vol. 970, no. 1, p. 95, July 2024.
doi: 10.3847/1538-4357/ad4a54.

\bibitem[Zacharias and Wagner(2016)]{2016A&A...588A.110Z}
M. Zacharias and S. J. Wagner,
The extended jet of AP Librae: Origin of the very high-energy $\gamma$-ray emission?,
\textit{A\&A}, vol. 588, p. A110, April 2016.
doi: 10.1051/0004-6361/201526698.

\end{thebibliography}



\end{document}